%% file: K0sKch_letterv11.tex
\documentclass[ALICE,manyauthors]{cernphprep}

\usepackage[comma,square,numbers,sort&compress]{natbib}
\usepackage{hyperref}
\usepackage{lineno}

\newcommand{\kzs}{\rm K^0_S}

\newcommand{\pvec}[1]{\vec{#1}\mkern2mu\vphantom{#1}}
\newcommand{\kstar}{k^*}
\newcommand{\vkstar}{\pvec{k}^*}
\newcommand{\rstar}{r^*}
\newcommand{\vrstar}{\pvec{r}^*}

\usepackage{hyperref}

\begin{document}%

\begin{titlepage}
\PHyear{2017}
\PHnumber{063}      
\PHdate{06 Apr}  
%

\title{Measuring K$^0_{\rm S}$K$^{\rm \pm}$ interactions \\
using Pb-Pb collisions at $\mathbf {\sqrt{s_{\rm NN}}=2.76}$ TeV}
\ShortTitle{K$^0_{\rm S}$K$^{\rm \pm}$ interaction using
Pb-Pb collisions}   

\Collaboration{ALICE Collaboration\thanks{See Appendix~\ref{app:collab} for the list of collaboration members}}
\ShortAuthor{ALICE Collaboration} 

\begin{abstract}
We present the first ever measurements of femtoscopic correlations between the K$^0_{\rm S}$ and K$^{\rm \pm}$ particles.
The analysis was performed on the data from Pb-Pb collisions at $\sqrt{s_{\rm NN}}=2.76$ TeV measured by the ALICE experiment.  The observed femtoscopic correlations are consistent with final-state interactions proceeding via the $a_0(980)$ resonance. 
The extracted kaon source radius and correlation strength parameters for K$^0_{\rm S}$K$^{\rm -}$ are found to be equal
within the experimental uncertainties to those for K$^0_{\rm S}$K$^{\rm +}$. Comparing the results of the present study with those
from published identical-kaon femtoscopic studies by ALICE, mass and coupling parameters for the $a_0$
resonance are tested. Our results are also compatible with the interpretation of 
the $a_0$ having a tetraquark structure instead of that of a diquark.
\end{abstract}
\end{titlepage}
\setcounter{page}{2}

\section{Introduction}
Identical boson femtoscopy, especially of
identical charged pions, has been used extensively over
the years to study experimentally the space-time geometry of the
collision region in high-energy particle and heavy-ion collisions
\cite{Lisa:2005dd}. 
Identical-kaon femtoscopy studies have also been carried out, recent examples of which are
the ones with Au-Au collisions at $\sqrt{s_{\rm NN}}=200$ GeV by the STAR Collaboration \cite{Abelev:2006gu} 
(K$^0_{\rm S}$K$^0_{\rm S}$) and with pp at $\sqrt{s}=7$ TeV and
Pb-Pb collisions at $\sqrt{s_{\rm NN}}=2.76$ TeV by the ALICE Collaboration \cite{Abelev:2012ms,Abelev:2012sq,Adam:2015vja} (K$^0_{\rm S}$K$^0_{\rm S}$ and K$^{\pm}$K$^{\pm}$).
The pair-wise interactions between the identical kaons that form the basis for femtoscopy are
for K$^{\rm \pm}$K$^{\rm \pm}$ quantum statistics and the Coulomb interaction, and for
K$^0_{\rm S}$K$^0_{\rm S}$ quantum statistics and the final-state interaction through the
$f_0(980)/a_0(980)$ threshold resonances.

One can also consider the case of non-identical kaon pairs, e.g. K$^0_{\rm S}$K$^{\rm \pm}$ pairs.
Besides the non-resonant channels which may be present, e.g.\ non-resonant elastic scattering
or free-streaming of the kaons from their freeze-out positions to the detector, 
the other only pair-wise interaction allowed for a K$^0_{\rm S}$K$^{\rm \pm}$ pair at freeze out from the collision system is a final-state interaction (FSI) through the $a_0(980)$ resonance. 
The other pair-wise interactions present for identical-kaon pairs are not present for K$^0_{\rm S}$K$^{\rm \pm}$ pairs because: a) there is no quantum statistics enhancement since the kaons are not identical, b) there is no Coulomb effect since one of the kaons is uncharged, and c) there is no strong FSI through the $f_0$ resonance since the kaon pair is in an $I=1$ isospin state, as is the $a_0$, whereas the $f_0$ is an $I=0$ state.

Another feature of the K$^0_{\rm S}$K$^{\rm \pm}$ FSI through the $a_0$ resonance is, due to the $a_0$ having strangeness $S=0$ and the K$^0_{\rm S}$ being a linear combination of the K$^0$ and ${\rm \overline{K}^0}$,
\begin{equation}
\left | {\rm K}^0_S \right \rangle=\frac{1}{\sqrt{2}}\left ( \left | {\rm K}^0 \right \rangle + \left | {\rm \overline{K}}^0 \right \rangle\right ),
\end{equation}
only the ${\rm \overline{K}^0}$K$^+$ pair from K$^0_{\rm S}$K$^{\rm +}$ and the K$^0$K$^-$
pair from K$^0_{\rm S}$K$^{\rm -}$ have $S=0$ and thus can form the $a_0$ resonance. This allows the possibility to study the K$^0$ and ${\rm \overline{K}^0}$ sources separately
since they are individually selected by studying K$^0_{\rm S}$K$^{\rm -}$ and
K$^0_{\rm S}$K$^{\rm +}$ pairs, respectively. An additional consequence of this feature is that only
$50\%$ of either the K$^0_{\rm S}$K$^{\rm -}$ or K$^0_{\rm S}$K$^{\rm +}$ detected pairs
will pass through the $a_0$ resonance. This is taken into account in the expression for the
model used to fit the correlation functions.

On the other hand, the natural requirement that the source sizes extracted from the K$^0_{\rm S}$K$^{\rm \pm}$ femtoscopy agree with those obtained for the K$^0_{\rm S}$K$^0_{\rm S}$ and K$^{\rm \pm}$K$^{\rm \pm}$ systems allows one to
study the
properties of the $a_0$ resonance itself. This is interesting in its own right since many studies discuss
the possibility that the $a_0$, listed by the Particle Data Group as a diquark light
unflavored meson state~\cite{Olive:2016xmw},
could be a four-quark state, i.e.
a tetraquark, or a ``${\rm \overline{K}}-$K molecule'' \cite{Martin1977,Antonelli2002,Achasov:2002ir,Achasov1,Santopinto:2006my,Jaffe:1976ig}. For example, the production cross section of the $a_0$ resonance in a reaction channel such as ${\rm K}^0{\rm K}^-\rightarrow a^-_0$ should depend on whether the $a^-_0$ is composed of ${\rm d\overline{u}}$ or ${\rm d\overline{s}s\overline{u}}$ quarks, the former requiring the annihilation of the ${\rm \overline{s}s}$ pair and the latter being a direct transfer of the quarks in the kaons to the $a^-_0$. The results from K$^0_{\rm S}$K$^-$ femtoscopy might be sensitive to these two different scenarios.

In this Letter, results from the first study of K$^0_{\rm S}$K$^{\rm \pm}$
femtoscopy are presented. This has been done for Pb-Pb collisions at $\sqrt{s_{\rm NN}}=2.76$ TeV measured by the ALICE experiment at the LHC \cite{Aamodt:2008zz}. 
The physics goals of the present K$^0_{\rm S}$K$^{\rm \pm}$ femtoscopy study are the following:
1) show to what extent the FSI through the $a_0$ resonance describes the correlation functions,
2) study the K$^0$ and ${\rm \overline{K^0}}$ sources to see if there are differences in the source parameters, and
3) test published $a_0$ mass and coupling parameters by comparisons with published
identical kaon results \cite{Adam:2015vja}.

\section{Description of experiment and data Selection}
The ALICE experiment and its performance in the LHC Run 1 $(2009-2013)$ are described in 
Ref.~\cite{Aamodt:2008zz} and Ref.~\cite{Abelev:2014ffa,Akindinov:2013tea}, respectively.
About $22\times 10^6$ Pb-Pb collision events with $0$--$10\%$ centrality class taken in 2011 were used in this analysis
(the average centrality in this range is 4.9\% due to a slight trigger inefficiency in the 8-10\% range).
Events were classified according to their centrality using the measured amplitudes in the 
V0 detectors, which consist of two arrays of scintillators located along the beamline and covering the full
azimuth
 \cite{Abelev:2013vea}. 
Charged particles were reconstructed and identified with the central barrel detectors located 
within a solenoid magnet with a field strength of $B=0.5$ T.
Charged particle tracking was performed using the 
Time Projection Chamber (TPC) \cite{Alme:2010ke} 
and the Inner Tracking System (ITS) \cite{Aamodt:2008zz}. The ITS allowed for high spatial resolution in determining the primary (collision) vertex.
Tracks were reconstructed and their momenta were obtained with the TPC.
A momentum resolution of less than 10 MeV/$c$ was typically obtained for the charged tracks
of interest in this analysis.
The primary vertex was obtained
from the ITS, the position of the primary vertex being constrained along the beam direction (the ``$z$-position'') to be within $\pm10$ cm of the center of the ALICE detector.
In addition to the standard track quality selections, the track selections based on the quality of track reconstruction fit and the number of detected tracking points in the TPC were used to ensure that only well-reconstructed tracks were taken in the analysis~\cite{Abelev:2014ffa,Akindinov:2013tea}.

Particle identification (PID) for reconstructed tracks was carried out using both the TPC and the Time-of-Flight (TOF) detector in the pseudorapidity range $|\eta| < 0.8$~\cite{Abelev:2014ffa,Akindinov:2013tea}.
For each PID method, a value was assigned to each track denoting the number of standard deviations between the measured track information and calculated values ($N_{\sigma}$) \cite{Adam:2015vja,Abelev:2014ffa,Akindinov:2013tea}.
For TPC PID, a parametrized Bethe-Bloch formula was used to calculate the specific energy 
loss $\left<{\rm d}E/{\rm d}x\right>$ in the detector expected for a particle with a given mass and momentum. For PID with TOF, the particle mass was used to calculate the expected time-of-flight as a function of track length and momentum. 
This procedure was repeated for four ``particle species hypotheses''---electron, pion, kaon and proton---, and, for each hypothesis, a different $N_{\sigma}$ value was obtained per detector.

\subsection{Kaon selection}
The methods used to select and identify individual K$^0_{\rm S}$ and K$^{\rm \pm}$ particles are the same as those used for the ALICE Pb-Pb K$^0_{\rm S}$K$^0_{\rm S}$
and K$^{\rm \pm}$K$^{\rm \pm}$ analyses \cite{Adam:2015vja}. These are now described below.

\subsubsection{K$^0_{\rm S}$ selection}
The K$^0_{\rm S}$ particles were reconstructed from the decay K$^0_{\rm S}\rightarrow\pi^+\pi^-$, with the daughter $\pi^+$ and $\pi^-$ tracks detected in the TPC and TOF detectors. 
Pions with $p_{\rm T}>0.15$ GeV/$c$ were accepted
(since for lower $p_{\rm T}$ track finding efficiency drops rapidly)
and the distance of closest approach to the primary vertex (DCA) of the reconstructed K$^0_{\rm S}$ was required to be less than 0.3 cm in all directions. 
The required $N_{\sigma}$ values for the pions were $N_{\sigma TPC} < 3$ and
$N_{\sigma TOF} < 3$ for $p>0.8$ GeV/$c$. An invariant mass distribution for the $\pi^+\pi^-$
pairs was produced and the K$^0_{\rm S}$ was defined to be resulting from a pair that fell
into the invariant mass range $0.480<m_{\pi^+\pi^-}<0.515$ GeV/$c^2$.

\subsubsection{K$^\pm$ selection}
Charged kaon tracks were also detected using the TPC and TOF detectors,
and were accepted if they were within the range 
$0.14<p_{\rm T}<1.5$ GeV/$c$.
In order to reduce the number of secondaries (for instance the charged particles produced in the detector material, particles from weak decays, etc.)
the primary charged kaon tracks were selected based on the DCA, such that the DCA
transverse to the beam direction was less than 2.4 cm and the DCA along the beam direction was
less than 3.2 cm. 
If the TOF signal were not available, 
the required $N_{\sigma}$ values for the charged kaons were $N_{\sigma TPC} < 2$ for $p_{\rm T}<0.5$ GeV/$c$, and the track was rejected for $p_{\rm T}>0.5$ GeV/$c$. If the TOF signal were also available and $p_{\rm T}>0.5$ GeV/$c$: $N_{\sigma TPC} < 3$
and $N_{\sigma TOF} < 2$ ($0.5<p_{\rm T}<0.8$ GeV/$c$), $N_{\sigma TOF} < 1.5$ ($0.8<p_{\rm T}<1.0$ GeV/$c$),
$N_{\sigma TOF} < 1$ ($1.0<p_{\rm T}<1.5$ GeV/$c$).

K$^0_{\rm S}$K$^{\rm \pm}$ experimental pair purity was estimated from a Monte Carlo (MC) study based on HIJING~\cite{Wang:1991hta} simulations using GEANT3~\cite{Brun:1994aa}  to model particle transport through the ALICE detectors. The purity
was determined from the fraction of the reconstructed MC simulated pairs that were identified as actual K$^0_{\rm S}$K$^{\rm \pm}$ pairs input from HIJING. The pair purity was
estimated to be 88\% for all kinematic regions studied in this analysis.

\section{Analysis methods}

\subsection{Experimental Correlation Functions}
This analysis studies the momentum correlations of K$^0_{\rm S}$K$^{\rm \pm}$ pairs using the two-particle correlation function, defined as 

\begin{equation}
C(k^*)=A(k^*)/B(k^*)
\end{equation}

where $A(k^*)$ is the measured distribution of pairs from the same event, $B(k^*)$ is the reference distribution of pairs from mixed events,
and $k^*$ is the magnitude of the momentum of each of the particles in the pair rest frame (PRF),

\begin{equation}
k^*=\sqrt{\frac{(s-m_{\rm K^0}^2-m_{\rm K^\pm}^2)^2-4m_{\rm K^0}^2m_{\rm K^\pm}^2}{4s}}
\end{equation}
where,
\begin{equation}
s=m_{\rm K^0}^2+m_{\rm K^\pm}^2+2E_{\rm K^0}E_{\rm K^\pm}-2\vec{p}_{\rm K^0}\cdot\vec{p}_{\rm K^\pm}
\end{equation}
and $m_{\rm K^0}$ ($E_{\rm K^0}$) and $m_{\rm K^\pm}$ ($E_{\rm K^\pm}$) are the rest masses
(total energies) of the K$^0_{\rm S}$ and K$^{\rm \pm}$, respectively.

The denominator $B(k^*)$ was formed by mixing K$^0_{\rm S}$ and K$^{\rm \pm}$ particles from each event with particles from ten other
events. The vertexes of the mixed events were constrained to be within 2 cm of each other in the $z$-direction. A centrality constraint on the mixed events was found not to be necessary for the narrow
centrality range, i.e.\ $0$--$10\%$, used in this analysis.
Correlation functions were obtained separately for two different magnetic field orientations
in the experiment and then
either averaged or fit separately, depending on the fitting method used (see below).

\begin{figure}[t!]
	\centering
		\includegraphics[scale=1.4]{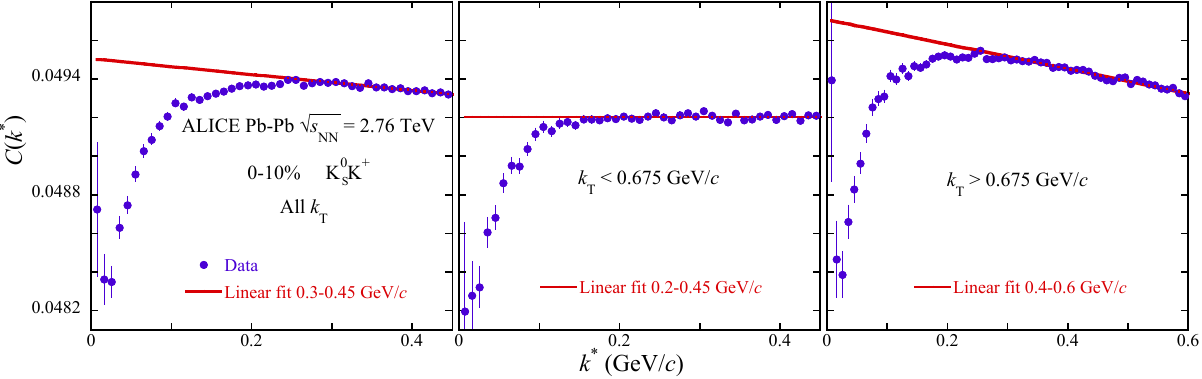}
	\caption{Examples of raw K$^0_{\rm S}$K$^{\rm +}$ correlation functions for the three $k_T$ bins with linear fits to the baseline at large $k^*$.  Statistical uncertainties are shown.}
	\label{fig1}
\end{figure}

Correlation functions were measured for three overlapping/non-exclusive
pair transverse momentum ($k_{\rm T} = |\textbf{p}_{\rm T,1}+\textbf{p}_{\rm T,2}|/2$)  bins: all $k_{\rm T}$, $k_{\rm T}<0.675$ and $k_{\rm T}>0.675$ GeV/$c$.
The mean $k_{\rm T}$ values for these three bins were 0.675, 0.425 and 0.970 GeV/$c$, respectively.
Figure~\ref{fig1} shows sample raw K$^0_{\rm S}$K$^{\rm +}$ correlation functions for these
three bins for one of the magnetic field orientations. One can see the main feature of the femtoscopic correlation function: the suppression due to the strong final-state interactions for small $k^*$. In the higher $k^*$ region, the effects of the $a_0$
appear to not be present and thus could be used as a reference, i.e.\ ``baseline'', for the $a_0$-based model fitted to $C(k^*)$ in order to extract the source parameters.
Also shown in the figure are linear fits to the baseline for large $k^*$. The effects on $C(k^*)$ by the $a_0$ resonance are mostly seen in the $k^*<0.2$ GeV/$c$ region, where the width of the $a_0$ region reflects the size of the kaon source (see equations below).

Correlation functions were corrected for momentum resolution effects using HIJING calculations.
HIJING was used to create two correlation functions: one in terms of the generator-level $k^*$ and one in terms of the simulated detector-level $k^*$. Because HIJING does not incorporate final-state interactions, weights were calculated using a 9th-order polynomial fit in $k^*$ to an experimental correlation function and were used when filling the same-event distributions. These weights were calculated using $k^*$. Then, the ratio of the ``ideal'' correlation function to the ``measured'' one (for each $k^*$ bin) was multiplied to the data correlation functions before the fit procedure. This correction mostly affected the lowest $k^*$ bins, increasing the extracted source parameters by several
percent.

\subsection{Final-state interaction model}
The K$^0_{\rm S}$K$^{\rm \pm}$ correlation functions were fit with functions that include a parameterization which incorporates strong FSI. It was assumed that the
FSI arises in the K$^0_{\rm S}$K$^{\rm \pm}$ channels due to the near-threshold resonance, $a_0$(980). This parameterization was introduced by R. Lednicky and is based on the model by R. Lednicky and V.L. Lyuboshitz~\cite{Lednicky:1981su,Lednicky:2005af} (see also Ref.~\cite{Abelev:2006gu} for more details on this parameterization).

Using an equal emission time approximation in the PRF~\cite{Lednicky:1981su}, the 
elastic K$^0_{\rm S}$K$^{\rm \pm}$ transition is written as a stationary solution $\Psi_{-\vkstar}(\vrstar)$ of the scattering problem in the PRF. The quantity
$\vrstar$ represents the emission separation of the pair in the PRF, and
the $-\vkstar$ subscript refers to a reversal of time from the emission process. At large distances this has the asymptotic form of a superposition of a plane wave and an outgoing spherical wave,
\begin{equation}
\Psi_{-\vkstar}(\vrstar) = e^{-i\vkstar \cdot \vrstar} + f(\kstar) \frac{e^{i\kstar\rstar}}{\rstar} \;,
\label{eq:FSIwave}
\end{equation}

where $f(k^*)$ is the $s$-wave K$^0$K$^-$ or $\overline{\rm K}^0$K$^+$ scattering amplitude whose contribution is the $s$-wave isovector $a_0$ resonance (see Eq.~11 in Ref.~\cite{Abelev:2006gu}), 

\begin{equation}
f(k^*) = \frac{\gamma_{a_0\rightarrow {\rm K\overline{K}}}}{m_{a_0}^2-s-i(\gamma_{a_0\rightarrow {\rm K\overline{K}}} k^*+\gamma_{a_0\rightarrow \pi\eta}k_{\pi\eta})}\;.
\label{eq:fit4}
\end{equation}

In Eq.~\ref{eq:fit4}, $m_{a_0}$ is the mass of the $a_0$ resonance, and $\gamma_{a_0\rightarrow {\rm K\overline{K}}}$ 
and $\gamma_{a_0\rightarrow \pi\eta}$ are the couplings of the $a_0$ resonance to the K$^0$K$^-$ (or $\overline{\rm K}^0$K$^+$) and $\pi\eta$ channels, respectively. Also, $s=4(m_{\rm K^0}^2+k^{*2})$ and $k_{\pi\eta}$ denotes the momentum in the second decay channel ($\pi\eta$) (see Table \ref{table1}).

The correlation function due to the FSI is then calculated by integrating $\Psi_{-\vkstar}(\vrstar)$ in the \textit{Koonin-Pratt equation}~\cite{Koonin:1977fh,Pratt:1990zq}
\begin{equation}
C(\vkstar) = \int {\rm d}^3 \, \vrstar \, S(\vrstar) \left| \Psi_{-\vkstar}(\vrstar) \right| ^2 \, ,
\label{eq:koonin}
\end{equation}
where $S(\vrstar)$ is a one-dimensional Gaussian source function of the PRF relative distance $\left| \vrstar \right| $ with a Gaussian width $R$ of the form
\begin{equation}
S(\vrstar) \sim e^{-\left| \vrstar \right| ^2/(4R^2)}\;.
\label{source}
\end{equation}

Equation~\ref{eq:koonin} can be integrated analytically for K$^0_{\rm S}$K$^{\rm \pm}$ correlations
with FSI for the one-dimensional case, with the result

\begin{equation}
C(k^*)=1+\lambda\alpha\left[\frac{1}{2}\left|\frac{f(k^*)}{R}\right|^2+\frac{2\mathcal{R}f(k^*)}{\sqrt{\pi}R}F_1(2k^* R)-\frac{\mathcal{I}f(k^*)}{R}F_2(2k^* R)\right],
\label{eq:fit2}
\end{equation}
where
\begin{equation}
F_1(z)\equiv\frac{\sqrt{\pi} e^{-z^2} \operatorname{erfi}(z)}{2 z};\qquad F_2(z)\equiv\frac{1-e^{-z^2}}{z}.
\label{eq:fit3}
\end{equation}
In the above equations $\alpha$ is the fraction of K$^{0}_{\rm S}$K$^{\pm}$ pairs that come from the K$^0$K$^-$ or  $\overline{\rm K}^0$K$^+$ system, set to 0.5 assuming symmetry in K$^0$ and $\overline{\rm K}^0$ production \cite{Abelev:2006gu}, $R$ is the radius parameter from the
spherical Gaussian source distribution given in Eq.~\ref{source}, and $\lambda$ is the correlation strength. The correlation strength is
unity in the ideal case of pure $a_0$-resonant FSI, perfect PID, a perfect Gaussian kaon source 
and the absence of long-lived resonances which decay into kaons.
Note that the form of the FSI term in
Eq.~\ref{eq:fit2} differs from the form of the FSI term for $\kzs\kzs$ correlations (Eq. 9 of Ref.~\cite{Abelev:2006gu}) by a factor of $1/2$ due to the non-identical particles in K$^0_{\rm S}$K$^{\rm \pm}$ correlations and thus the absence of the requirement to symmetrize the wavefunction given
in Eq.~\ref{eq:FSIwave}.

As seen in Eq.~\ref{eq:fit4}, the K$^0$K$^-$ or $\overline{\rm K}^0$K$^+$ s-wave scattering amplitude depends on the $a_0$ mass and decay couplings. In the present work, we have
taken the values used in Ref.~\cite{Abelev:2006gu}
which have been extracted from the analysis of the $a_0\rightarrow\pi\eta$ spectra of several 
experiments \cite{Martin1977,Antonelli2002,Achasov1,Achasov:2002ir}, shown in Table \ref{table1}.
The extracted $a_0$ mass and decay couplings have a range of values for
the various references. Except for the Martin reference~\cite{Martin1977}, which extracts the $a_0$ values
from the reaction 4.2 GeV/$c$ incident momentum K$^-+p\rightarrow\Sigma^+(1385)\pi^-\eta$
using a two-channel Breit-Wigner formula, the other references
extract the $a_0$ values from the radiative $\phi$-decay data, i.e.\ $\phi\rightarrow\pi^0\eta\gamma$,
from the KLOE collaboration~\cite{Aloisio:2002bsa}. These latter three references apply a model that
assumes, after taking into account the $\phi\rightarrow\pi^0\rho^0\rightarrow\pi^0\eta\gamma$
background process, that the $\phi$ decays to the $\pi^0\eta\gamma$ final
state through
the intermediate processes $\phi\rightarrow {\rm K}^+{\rm K}^-\gamma\rightarrow a_0\gamma$ or
$\phi\rightarrow {\rm K}^+{\rm K}^-\rightarrow a_0\gamma$, 
i.e.\ the ``charged kaon loop model''~\cite{Achasov:2002ir}. The main difference between these
analyses is that the Antonelli reference~\cite{Antonelli2002} assumes a fixed
$a_0$ mass in the fit of this model to the $\pi^0\eta$ data, whereas the Achasov1 and
Achasov2 analyses~\cite{Achasov:2002ir} allow the $a_0$ mass to be a free parameter in the two different
fits made to the data.
It is assumed in the present analysis
that these decay couplings will also be valid for K$^0$K$^-$ and $\overline{\rm K}^0$K$^+$
scattering due to isospin invariance.
Correlation functions were fitted with all four of these cases to see the effect
on the extracted source parameters.

\begin{table}
 \centering
 \begin{tabular}{| c | c | c | c |}
  \hline
  Reference & $m_{a_0}$ & $\gamma_{a_0K\bar{K}}$ & $\gamma_{a_0\pi\eta}$ \\ \hline
  Martin~\cite{Martin1977} & 0.974 & 0.333 & 0.222 \\ \hline
  Antonelli~\cite{Antonelli2002} & 0.985 & 0.4038 & 0.3711 \\ \hline
  Achasov1~\cite{Achasov:2002ir} & 0.992 & 0.5555 & 0.4401 \\ \hline
  Achasov2~\cite{Achasov:2002ir} & 1.003 & 0.8365 & 0.4580 \\  
  \hline
  \end{tabular}
  \caption{The $a_0$ masses and coupling parameters, all in GeV (taken from Ref.\cite{Abelev:2006gu}).}
  \label{table1}
\end{table}

\subsection{Fitting methods}
In order to estimate the systematic errors in the fitting method used to extract 
$R$ and $\lambda$ using Eq.~\ref{eq:fit2},
two different methods, judged to be equally valid, have been used to handle the effects of the baseline: 1) a separate linear
fit to the ``baseline region,'' followed by fitting Eq.~\ref{eq:fit2} to the correlation function divided by
the linear fit to extract the source parameters, 
and 2) a combined fit of Eq.~\ref{eq:fit2} and a quadratic function describing the baseline where
the source parameters and the parameters of the quadratic function are fitted simultaneously.
The source parameters are extracted for each case from both methods and averaged, the 
symmetric systematic error for each case due to the fitting method being one-half of the difference between the two methods. Both fitting methods will now be described in more detail.

\subsubsection{Linear baseline method} 
In the ``linear baseline method,'' for the all $k_{\rm T}$, $k_{\rm T}<0.675$ and $k_{\rm T}>0.675$ GeV/$c$ bins the $a_0$
regions were taken to be $k^*<0.3$, $k^*<0.2$ and $k^*<0.4$ GeV/$c$, respectively.
In the higher $k^*$ region it was assumed that effects of the $a_0$ were not present and thus can be
used as a reference, i.e.\ ``baseline'', for the $a_0$-based model fitted to $C(k^*)$, which 
was averaged over the two magnetic field orientations used in the experiment,
to extract the source parameters.
For the three $k_{\rm T}$ bins, linear fits were made in the
$k^*$ ranges $0.3$--$0.45$, $0.2$--$0.45$ and $0.4$--$0.6$ GeV/$c$, respectively, and the correlation functions
were divided by these fits to remove baseline effects extending into the low-$k^*$ region. 
These ranges were taken to define the baselines since the measured correlation functions 
were found to be linear here. For larger values of $k^*$ the correlation functions became non-linear.
The baseline was studied
using HIJING MC calculations which take into account the detector characteristics as described
earlier.  The $C(k^*)$ distributions obtained from HIJING do not show
suppressions at low $k^*$ as seen in Fig.~\ref{fig1} but rather show linear distributions over the
entire ranges in $k^*$ shown in the figure. HIJING also shows the baseline becoming non-linear for
larger values of $k^*$, as seen in the measurements. The MC generator code 
AMPT~\cite{Lin:2004en} was also used to study the baseline. AMPT is similar to HIJING but also
includes final-state rescattering effects. AMPT calculations also showed linear baselines in the
$k^*$ ranges used in the present analysis, becoming non-linear for larger $k^*$. Both HIJING and
AMPT qualitatively show the same direction of changes in the slopes of the baseline vs. $k_{\rm T}$
as seen in the data, but AMPT more accurately described the slope values themselves, 
suggesting that final-state rescattering plays a role in the $k_{\rm T}$ dependence of the 
baseline slope.
The systematic uncertainties on the extracted source parameters due to the assumption of
linearity in these $k^*$ regions were estimated from HIJING to be less than 1\%.

\begin{figure}[t!]
	\centering
		\includegraphics[scale=1.4]{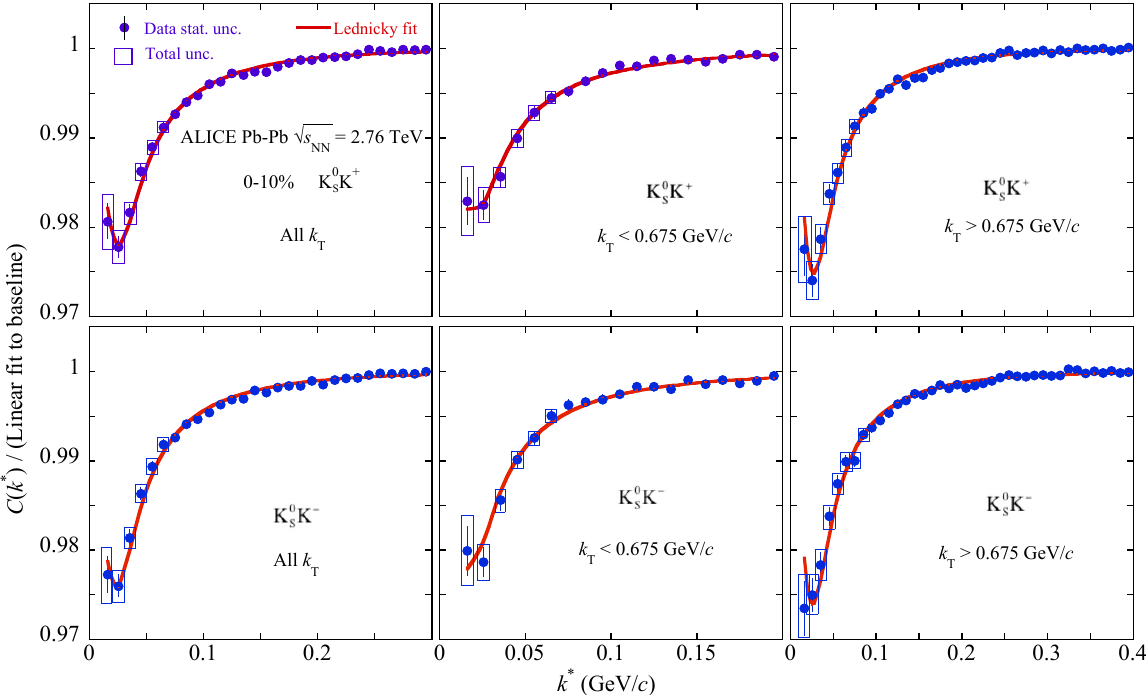}
	\caption{Examples of K$^0_{\rm S}$K$^+$ and K$^0_{\rm S}$K$^-$ correlation functions
	divided by linear fits to the baseline
	with the Lednicky parameterization using the Achasov2 \cite{Achasov:2002ir}
parameters. Statistical (lines) and the linear sum of statistical and systematic uncertainties (boxes) are shown.}
	\label{fig2}
\end{figure}

Figure~\ref{fig2} shows examples of K$^0_{\rm S}$K$^+$ and K$^0_{\rm S}$K$^-$ correlation functions
divided by linear fits to the baseline with Eq.~\ref{eq:fit2} using the Achasov2
parameters.
One can see the main feature of the femtoscopic correlation function: the suppression due to the strong final-state interactions for small $k^*$.
As seen, the $a_0$ FSI parameterization gives an excellent representation of the ``signal region''
of the data,
i.e.\ the suppression of the correlation functions in the $k^*$ range 0 to about 0.15 GeV/$c$.

\subsubsection{Quadratic baseline method}
In the ``quadratic baseline method,'' $R$ and $\lambda$ are extracted
assuming a quadratic baseline function by fitting the product of a quadratic function and the
Lednicky equation, Eq.~\ref{eq:fit2}, to the raw correlation functions for each of the two magnetic
field orientations used in the experiment, such as shown in 
Fig.~\ref{fig1}, i.e.\ ,

\begin{equation}
C_{raw}^{fit}(k^*)=a(1-bk^*+ck^{*2})C(k^*)
\label{eq:quad}
\end{equation}

where $C(k^*)$ is given by Eq.~\ref{eq:fit2}, and $a$, $b$ and $c$ are fit parameters. 
Eq.~\ref{eq:quad} is fit to the same $k^*$ ranges as shown in Fig.~\ref{fig1}, i.e.
$0$--$0.45$ GeV/$c$ for all $k_{\rm T}$ and $k_{\rm T}<0.675$ GeV/$c$, and $0$--$0.6$ GeV/$c$ for
$k_{\rm T}>0.675$ GeV/$c$. The fits to the experimental correlation functions
are found to be of similar good quality as seen for the linear baseline method fits shown
in Fig.~\ref{fig2}.

\subsection{Systematic uncertainties}
Systematic uncertainties on the extracted source parameters were estimated by varying the ranges of kinematic and PID cut values on the
data by $\pm10\%$ and $\pm20\%$, as well as from MC simulations. The main systematic
uncertainties on the extracted values of $R$ and $\lambda$ due to various sources, not including the
baseline fitting method, are: a) $k^*$ fitting range: 2\%, b) single-particle and pair cuts (e.g. DCA cuts,
PID cuts, pair separation cuts): $2\%$--$4\%$ for $R$ and $3\%$--$8\%$ for $\lambda$, and 
c) pair purity: 1\% on $\lambda$.
Combining the individual systematic
uncertainties in quadrature, the total systematic uncertainties on the extracted source parameters,
not including the baseline fitting method contribution,
are in the ranges $3\%$--$5\%$ for $R$ and $4\%$--$8\%$ for $\lambda$.

As mentioned earlier, for the two fitting methods, the source parameters are extracted for each case from both methods and averaged, the 
symmetric systematic error for each case due to the fitting method being one-half of the difference between the two methods.
The baseline fitting method systematic error thus obtained is added in quadrature with the systematic errors
given above. It is found that the size of the baseline fitting method systematic
errors are about 50\% larger for $R$ and of similar magnitude for $\lambda$ as those quoted
above for the non-fitting-method systematic errors.

\section{Results and discussion}
\begin{figure}[t!]
	\centering
		\includegraphics[scale=1.]{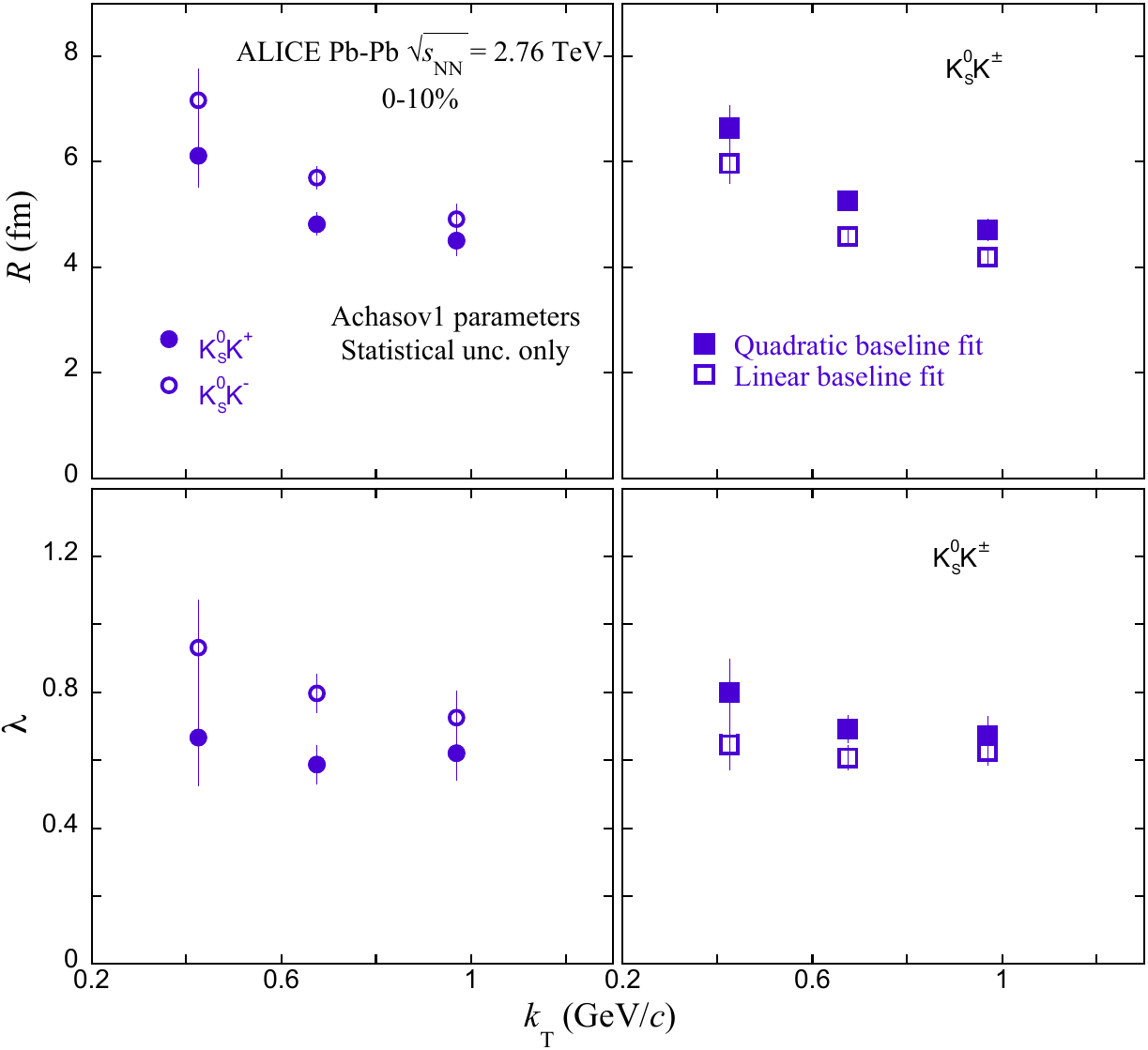}
	\caption{Sample results for the $R$ and $\lambda$ parameters extracted in the 
	present analysis from K$^0_{\rm S}$K$^{\rm \pm}$ femtoscopy using the Achasov1
	parameters. The left column compares K$^0_{\rm S}$K$^{\rm +}$
and K$^0_{\rm S}$K$^{\rm -}$ results from the quadratic baseline fit method, and the right column
compares results averaged over K$^0_{\rm S}$K$^{\rm +}$
and K$^0_{\rm S}$K$^{\rm -}$ for the quadratic baseline fits and the linear baseline fits.
Statistical uncertainties are plotted for all results.}
	\label{fig3}
\end{figure}

Figure~\ref{fig3} shows sample results for the $R$ and $\lambda$ parameters extracted in the 
present analysis from K$^0_{\rm S}$K$^{\rm \pm}$ femtoscopy using the Achasov1
parameters. The left column compares K$^0_{\rm S}$K$^{\rm +}$
and K$^0_{\rm S}$K$^{\rm -}$ results from the quadratic baseline fit method, and the right column
compares results averaged over K$^0_{\rm S}$K$^{\rm +}$
and K$^0_{\rm S}$K$^{\rm -}$ for the quadratic baseline fits and the linear baseline fits.
As it is usually the case in femtoscopic analyses, the fitted $R$ and $\lambda$ parameters are
correlated. The fitting (statistical) uncertainties are taken to be the extreme
values of the $1\sigma$ fit contours in $R$ vs. $\lambda$.
Statistical uncertainties are plotted for all results. It is seen in the figure
that the $R$ and $\lambda$ values for K$^0_{\rm S}$K$^{\rm -}$ have a slight tendency to be larger than those for  K$^0_{\rm S}$K$^{\rm +}$. Such a difference could result from the K$^{\rm -}$ -- nucleon scattering cross section being larger than that for K$^{\rm +}$ -- nucleon (see Fig.~51.9 of Ref.~\cite{Olive:2016xmw}), 
possibly resulting in more
final-state rescattering for the K$^{\rm -}$.
Since the difference is not significant once systematic
uncertainties are taken into account, K$^0_{\rm S}$K$^{\rm +}$
and K$^0_{\rm S}$K$^{\rm -}$ are averaged over in the final results. The difference in the extracted parameters between the two baseline fitting methods is also seen to be small, and is accounted for as a systematic error, as described earlier.

\begin{figure}[t!]
	\centering
		\includegraphics[scale=1.]{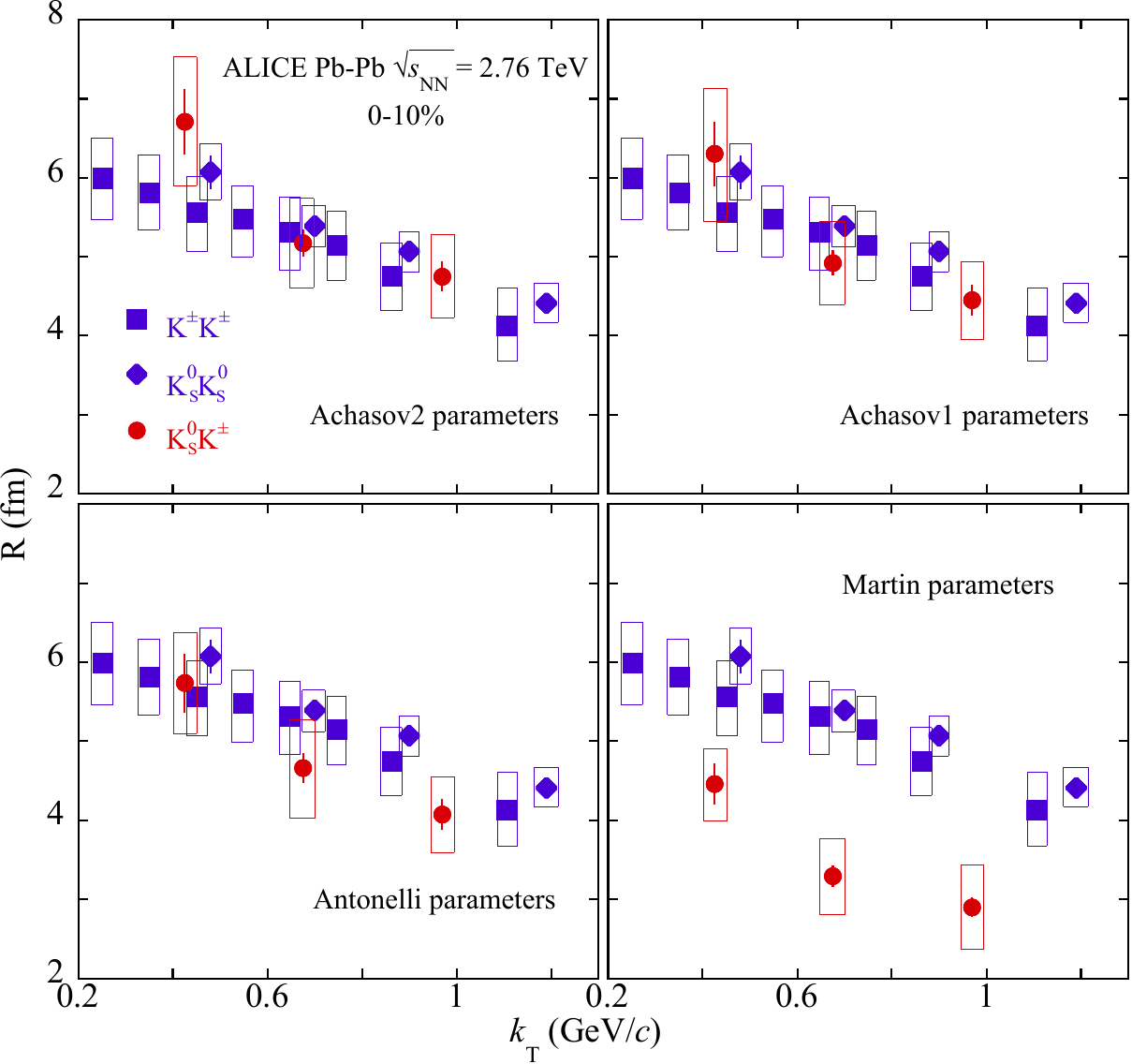}
	\caption{Source radius parameter, $R$, extracted in the present analysis from 
K$^0_{\rm S}$K$^{\rm \pm}$ femtoscopy averaged over K$^0_{\rm S}$K$^{\rm +}$
and K$^0_{\rm S}$K$^{\rm -}$ and the two baseline fit methods (red symbols),
along with
comparisons with identical kaon results from ALICE~\cite{Adam:2015vja} (blue symbols).
Statistical (lines) and the linear sum of statistical and systematic uncertainties (boxes) are shown.}
	\label{fig4}
\end{figure}

\begin{figure}[t!]
	\centering
		\includegraphics[scale=1.]{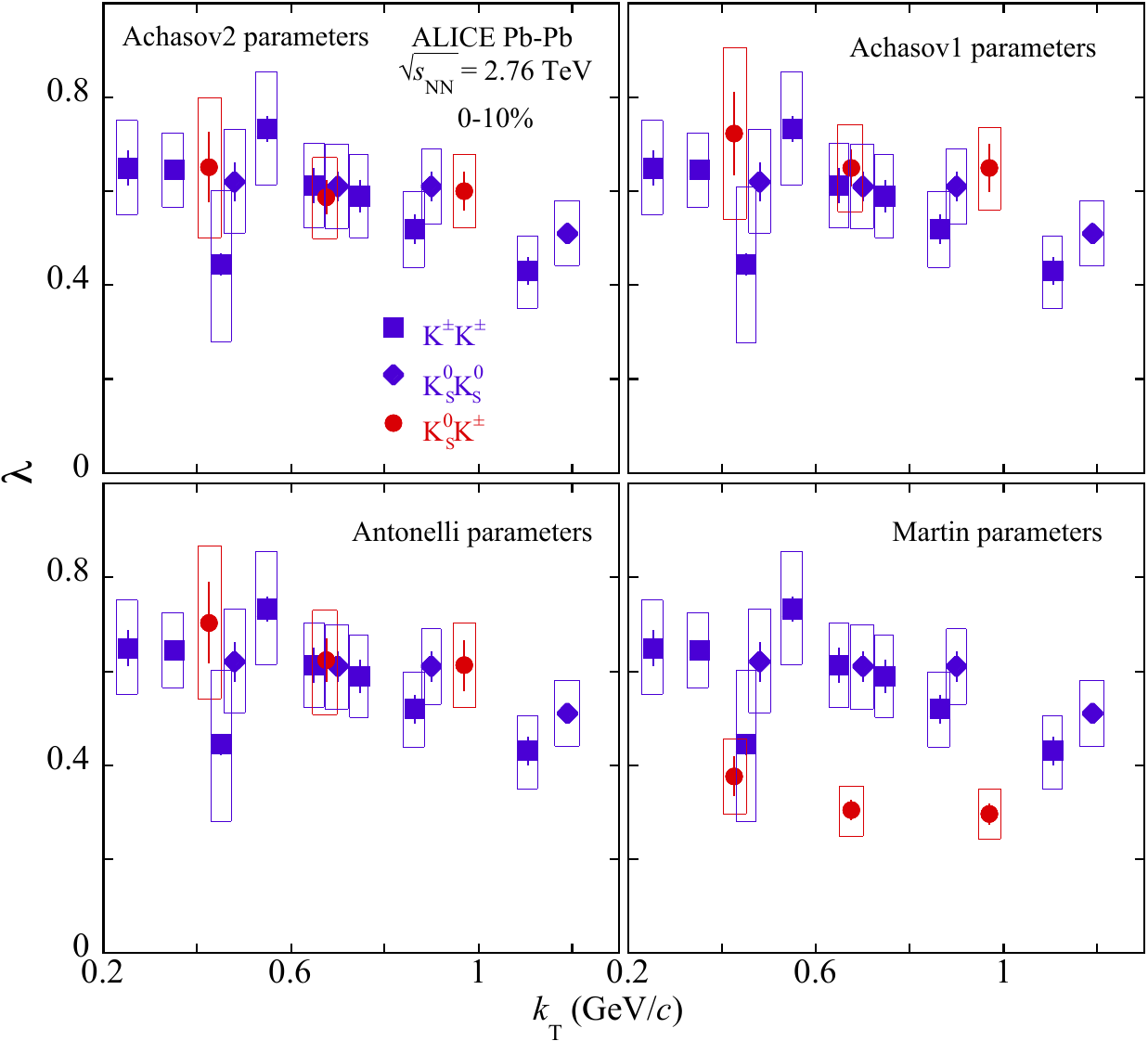}
	\caption{Correlation strength parameter, $\lambda$, extracted in the present analysis from 
K$^0_{\rm S}$K$^{\rm \pm}$ femtoscopy averaged over K$^0_{\rm S}$K$^{\rm +}$
and K$^0_{\rm S}$K$^{\rm -}$ and the two baseline fit methods (red symbols), along with
comparisons with identical kaon results from ALICE~\cite{Adam:2015vja} (blue symbols).
Statistical (lines) and the linear sum of statistical and systematic uncertainties (boxes) are shown.}
	\label{fig5}
\end{figure}

The results for the $R$ and $\lambda$ parameters extracted in the present analysis from 
K$^0_{\rm S}$K$^{\rm \pm}$ femtoscopy, averaged over the two baseline fit methods and
averaged over K$^0_{\rm S}$K$^{\rm +}$
and K$^0_{\rm S}$K$^{\rm -}$, are presented in Table~\ref{tab:results} and in Figs. \ref{fig4}
and \ref{fig5}. Fit results are shown for all four parameter sets given in Table \ref{table1}.
Figs.~\ref{fig4} and \ref{fig5} also show
comparisons with identical kaon results for the same
collision system and energy from ALICE from Ref.~\cite{Adam:2015vja}.
Statistical and total uncertainties are shown for all results.

\begin{table}
 \centering
 \begin{tabular}{| c | c | c | c | c |}
  \hline
  Parameters & $R$ (fm) or $\lambda$  & all $k_{\rm T}$ & $k_{\rm T}<0.675$ GeV/$c$ & $k_{\rm T}>0.675$ GeV/$c$  \\ \hline
  Achasov2 & $R$ & $5.17\pm0.16\pm0.41$  & $6.71\pm0.40\pm0.42$ & $4.75\pm0.18\pm0.36$  \\ \hline
        & $\lambda$ & $0.587\pm0.034\pm0.051$ & $0.651\pm0.073\pm0.076$ & $0.600\pm0.040\pm0.034$ \\ \hline
  Achasov1 & $R$ & $4.92\pm0.15\pm0.39$ & $6.30\pm0.40\pm0.43$ & $4.49\pm0.18\pm0.30$ \\ \hline
        & $\lambda$ & $0.650\pm0.038\pm0.056$ & $0.723\pm0.087\pm0.091$ & $0.649\pm0.048\pm0.038$ \\ \hline  
  Antonelli & $R$ & $4.66\pm0.17\pm0.46$ & $5.74\pm0.36\pm0.26$ & $4.07\pm0.18\pm0.29$ \\ \hline
        & $\lambda$ & $0.624\pm0.044\pm0.058$ & $0.703\pm0.085\pm0.077$ & $0.613\pm0.052\pm0.037$ \\ \hline  
  Martin & $R$ & $3.29\pm0.12\pm0.35$ & $4.46\pm0.25\pm0.20$ & $2.90\pm0.11\pm0.41$ \\ \hline
        & $\lambda$ & $0.305\pm0.020\pm0.033$ & $0.376\pm0.041\pm0.037$ & $0.296\pm0.021\pm0.030$ \\  
  \hline
  \end{tabular}
  \caption{Fit results for $R$ and $\lambda$ extracted in the present analysis from 
K$^0_{\rm S}$K$^{\rm \pm}$ femtoscopy averaged over K$^0_{\rm S}$K$^{\rm +}$
and K$^0_{\rm S}$K$^{\rm -}$. Statistical and systematic errors are also shown.}
  \label{tab:results}
\end{table}

As shown in Fig.~\ref{fig4}, both Achasov parameter sets, with the larger $a_0$ masses and decay couplings, appear to give $R$ values that agree best with those obtained from identical-kaon femtoscopy. The Antonelli parameter set appears to give slightly lower values.
Comparing the measured $R$ values between
K$^0_{\rm S}$K$^0_{\rm S}$ and K$^{\pm}$K$^{\pm}$
in Fig.~\ref{fig4} they are seen to agree with each other within the uncertainties. 
In fact, the only reason for the femtoscopic K$^0_{\rm S}$K$^{\rm \pm}$ radii to be different from the
K$^0_{\rm S}$K$^0_{\rm S}$ and K$^{\pm}$K$^{\pm}$ ones would be if the K$^0_{\rm S}$ and 
K$^{\pm}$ sources were displaced with respect to each other. This is not expected because the collision dynamics is governed by strong interactions for which the isospin symmetry applies.

The results for the correlation strength parameters $\lambda$ are shown in Fig.~\ref{fig5}.
The $\lambda$ parameters from K$^0_{\rm S}$K$^{\rm \pm}$ and K$^{\pm}$K$^{\pm}$
are corrected for experimental purity~\cite{Adam:2015vja}.The K$^0_{\rm S}$K$^0_{\rm S}$ 
pairs have a
high purity of $> 90\%$, so the corresponding correction was 
neglected~\cite{Adam:2015vja}(see the earlier discussion on purity).
Statistical and total uncertainties are shown for all results.

The K$^0_{\rm S}$K$^{\rm \pm}$ $\lambda$ values, with the exception
of the Martin parameters, appear to be in agreement with the $\lambda$ values for the identical kaons. All of the $\lambda$ values are seen to be measured to be about 0.6, i.e. less than the ideal value of unity,
which can be due to the contribution of kaons from K$^*$ decay 
($\Gamma\sim50$ MeV, where $\Gamma$ is the decay width)  
and from other long-lived resonances (such as the $D$-meson) distorting the
spatial kaon source distribution away from the ideal Gaussian which is 
assumed in the fit function~\cite{Humanic:2013xga}. One would
expect that the K$^0_{\rm S}$K$^{\rm \pm}$ $\lambda$ values agree with those
from the identical kaons if the FSI for the K$^0_{\rm S}$K$^{\rm \pm}$ went solely
through the $a_0$ resonant channel since this analysis should see the same source
distribution.

In order to obtain a more quantitative comparison of the present results for $R$ and $\lambda$
with the identical kaon results, the $\chi^2/{\rm ndf}$ is calculated for $R$ and $\lambda$ for each parameter
set,

\begin{equation}
\chi_\omega^2/{\rm ndf} = \frac{1}{\nu}\sum_{i=1}^{3}\frac{[\omega_i(K^0_{\rm S}K^{\rm \pm})-\omega_i(KK)]^2}{\sigma_i^2}
\label{chi2}
\end{equation}

where $\omega$ is either $R$ or $\lambda$, $i$ runs over the three $k_T$ values, the number
of degrees of freedom taken is
${\rm ndf}=3$ and $\sigma_i$ is the sum of the statistical and systematic uncertainties on the 
$i^{th}$ K$^0_{\rm S}$K$^{\rm \pm}$ extracted parameter
(Note that the all $k_{\rm T}$ bin indeed contains the kaon pairs that make up the $k_{\rm T}<0.675$ GeV/$c$ and $k_{\rm T}>0.675$ GeV/$c$ bins, but in addition it contains an equal number of new pair combinations between the kaons in the $k_{\rm T}<0.675$ GeV/$c$ and $k_{\rm T}>0.675$ GeV/$c$ bins. So for the purposes of this simple comparison, we approximate the all $k_{\rm T}$ bin as being independent).
The linear sum of the statistical and systematic uncertainties is used for $\sigma_i$ to be consistent with the linear sum of the statistical and systematic uncertainties plotted on 
the points in Figs.~\ref{fig4} and ~\ref{fig5}.
The quantity
$\omega_i(KK)$ is determined by fitting a quadratic to the identical kaon results and evaluating
the fit at the average $k_T$ values of the K$^0_{\rm S}$K$^{\rm \pm}$ measurements.
Table~\ref{tab:chisq} summarizes the results for each parameter set and the extracted
p-values. 
As seen, the Achasov2, Achasov1 and Antonelli parameter sets are consistent with the
identical kaon results for  both $R$
and $\lambda$. 
The Martin parameter set is seen to have vanishingly small p-values for both $R$ and $\lambda$ and is thus in clear disagreement with the identical kaon results, as can easily be seen by examining Figs.~\ref{fig4} and \ref{fig5}.

In order to quantitatively estimate the size of the non-resonant channel present, the ratio
$\left \langle \frac{\lambda(K^0_{\rm S}K^{\rm \pm})}{\lambda(KK)} \right \rangle$ has been
calculated for each parameters set, where the average is over the three $k_T$ values
and the uncertainty is calculated from the average of the statistical+systematic 
uncertainties on the K$^0_{\rm S}$K$^{\rm \pm}$ parameters. These values are shown in
the last column of Table~\ref{tab:chisq}. Disregarding the Martin value, the smallest
value this ratio can take within the uncertainties is 0.87 (from the Achasov2 paramters)
which would thus allow at most a 13\% non-resonant contribution.

\begin{table}
 \centering
 \begin{tabular}{| c | c | c | c | c | c |}
  \hline
  Parameters & $\chi_R^2/{\rm ndf}$ & $R$ p-value & $\chi_\lambda^2/{\rm ndf}$ & $\lambda$ p-value & $\left \langle \frac{\lambda(K^0_{\rm S}K^{\rm \pm})}{\lambda(KK)} \right \rangle$ \\ \hline
  Achasov2 & 0.456 & 0.713 & 0.248 & 0.863 & 1.04$\pm$0.17  \\ \hline
  Achasov1 & 0.583 & 0.626 & 0.712 & 0.545 & 1.14$\pm$0.20 \\ \hline
  Antonelli & 1.297 & 0.273 & 0.302 & 0.824 & 1.09$\pm$0.20 \\ \hline
  Martin & 14.0 & 0.000 & 22.2 & 0.000 & 0.55$\pm$0.10 \\  
  \hline
  \end{tabular}
  \caption{Comparisons of $R$ and $\lambda$ from K$^0_{\rm S}$K$^{\rm \pm}$ with identical kaon results.}
  \label{tab:chisq}
\end{table}

The results of this study presented above clearly show that the measured 
K$^0_{\rm S}$K$^{\rm \pm}$ have dominantly undergone a FSI through the $a_0$ resonance.
This is remarkable considering that we measure in Pb-Pb collisions the average separation between the two kaons at freeze out to be $\sim 5$ fm, and 
due to the short-ranged nature of the strong interaction of $\sim 1$ fm
this would seem to not encourage a FSI but rather encourage free-streaming of the kaons to the detector resulting in a ``flat'' correlation function.
A dominant FSI is what might be expected if the $a_0$ would be a four-quark, i.e.\ tetraquark, state
or a ``${\rm \overline{K}}-$K molecule.''
There appears to be no calculations in the literature for the tetraquark vs. diquark production cross sections
for the interaction ${\rm K\overline{K}} \rightarrow a_0$, but qualitative arguments compatible with
the $a_0$ being a four--quark state can be made based on the present measurements.
The main argument
in favor of this is that the reaction channel
${\rm K}^0{\rm K}^-\rightarrow a^-_0$ ($\overline{\rm K}^0$K$^+\rightarrow a^+_0$) is strongly favored if the $a^-_0$ ($a^+_0$) is composed of ${\rm d\overline{s}s\overline{u}}$ (${\rm \overline{d}s\overline{s}u}$) quarks such that a direct transfer of the quarks in the kaons to the $a^-_0$ ($a^+_0$) has taken
place, since this is an ``OZI superallowed''  reaction~\cite{Jaffe:1976ig}. 
The ``OZI rule'' can be stated as
``an inhibition associated with the creation or annihilation of quark lines''~\cite{Jaffe:1976ig}.
Thus, a diquark $a_0$ final state is less favored according to the OZI rule since it would require the annihilation
of the strange quarks in the kaon interaction. This would allow for the possibility of a
significant non-resonant or free-streaming channel for the kaon interaction that would 
result in a $\lambda$ 
value below
the identical-kaon value by diluting the $a_0$ signal.
As mentioned above, the collision geometry itself also suppresses the annihilation of the 
strange quarks due to the
large separation between the kaons at freeze out.
Note that this assumes that the $C(k^*)$ distribution of a non-resonant channel would be mostly ``flat''
or ``monotonic'' in shape and not showing a strong resonant-like signal
as seen for the $a_0$ in Fig.~\ref{fig1} and Fig.~\ref{fig2}. This assumption is clearly true in
the free-streaming case, which is assumed in Eq.~\ref{eq:fit2} in setting $\alpha=0.5$ due to the
non-resonant kaon combinations.
A similar argument, namely that the success of the ``charged kaon loop
model'' in describing the radiative $\phi$-decay data favors the $a_0$ as a tetraquark state,
is given in Ref.~\cite{Achasov:2002ir}.

\section{Summary}
In summary, femtoscopic correlations with K$^0_{\rm S}$K$^{\rm \pm}$ pairs have been studied for the first time. This new femtoscopic method was applied to data from central Pb-Pb collisions at 
$\sqrt{s_{\rm NN}}=2.76$ TeV by the LHC ALICE experiment. Correlations in the K$^0_{\rm S}$K$^{\rm \pm}$ pairs are produced by final-state interactions which proceed through the a$_{\rm 0}$(980) resonance. The $a_0$ resonant FSI is seen to give an excellent representation of the shape of the signal region in the present study. The differences between 
${\rm \overline{K}^0}$K$^+$ and K$^0$K$^-$
for the extracted $R$ and $\lambda$ values are found to be insignificant within the uncertainties of the present study. The three larger $a_0$ mass and decay parameter sets are favored by the comparison with the
identical kaon results. The present results are also compatible with the interpretation of the $a_0$ resonance
as a tetraquark state.
This work should provide a constraint on models that are used to predict kaon-kaon interactions \cite{Oller:1998hw,HongXiem:2014tda}.
It will be interesting to apply
K$^0_{\rm S}$K$^{\rm \pm}$ femtoscopy to other collision energies, e.g.\ the higher LHC energies
now available,  and bombarding species, e.g.\ proton-proton collisions, since the different source sizes encountered in these cases will probe the interaction of the K$^0_{\rm S}$ with the K$^{\rm \pm}$ in
different sensitivity ranges (i.e.\ see the $R$ dependence in Eq.~\ref{eq:fit2}).

\newenvironment{acknowledgement}{\relax}{\relax}
\begin{acknowledgement}
\section*{Acknowledgements}
\input{fa_2017-04-05.tex}    
\end{acknowledgement}

\bibliographystyle{utphys}   
\bibliography{K0sKch_lettervx.bib}

\newpage
\appendix
\section{The ALICE Collaboration}
\label{app:collab}
\input{Alice_Authorlist_2017-Apr-05.tex}  
\end{document}

%% file: fa_2017-04-05.tex

The ALICE Collaboration would like to thank all its engineers and technicians for their invaluable contributions to the construction of the experiment and the CERN accelerator teams for the outstanding performance of the LHC complex.
The ALICE Collaboration gratefully acknowledges the resources and support provided by all Grid centres and the Worldwide LHC Computing Grid (WLCG) collaboration.
The ALICE Collaboration acknowledges the following funding agencies for their support in building and running the ALICE detector:
A. I. Alikhanyan National Science Laboratory (Yerevan Physics Institute) Foundation (ANSL), State Committee of Science and World Federation of Scientists (WFS), Armenia;
Austrian Academy of Sciences and Nationalstiftung f\"{u}r Forschung, Technologie und Entwicklung, Austria;
Ministry of Communications and High Technologies, National Nuclear Research Center, Azerbaijan;
Conselho Nacional de Desenvolvimento Cient\'{\i}fico e Tecnol\'{o}gico (CNPq), Universidade Federal do Rio Grande do Sul (UFRGS), Financiadora de Estudos e Projetos (Finep) and Funda\c{c}\~{a}o de Amparo \`{a} Pesquisa do Estado de S\~{a}o Paulo (FAPESP), Brazil;
Ministry of Science \& Technology of China (MSTC), National Natural Science Foundation of China (NSFC) and Ministry of Education of China (MOEC) , China;
Ministry of Science, Education and Sport and Croatian Science Foundation, Croatia;
Ministry of Education, Youth and Sports of the Czech Republic, Czech Republic;
The Danish Council for Independent Research | Natural Sciences, the Carlsberg Foundation and Danish National Research Foundation (DNRF), Denmark;
Helsinki Institute of Physics (HIP), Finland;
Commissariat \`{a} l'Energie Atomique (CEA) and Institut National de Physique Nucl\'{e}aire et de Physique des Particules (IN2P3) and Centre National de la Recherche Scientifique (CNRS), France;
Bundesministerium f\"{u}r Bildung, Wissenschaft, Forschung und Technologie (BMBF) and GSI Helmholtzzentrum f\"{u}r Schwerionenforschung GmbH, Germany;
General Secretariat for Research and Technology, Ministry of Education, Research and Religions, Greece;
National Research, Development and Innovation Office, Hungary;
Department of Atomic Energy Government of India (DAE) and Council of Scientific and Industrial Research (CSIR), New Delhi, India;
Indonesian Institute of Science, Indonesia;
Centro Fermi - Museo Storico della Fisica e Centro Studi e Ricerche Enrico Fermi and Istituto Nazionale di Fisica Nucleare (INFN), Italy;
Institute for Innovative Science and Technology , Nagasaki Institute of Applied Science (IIST), Japan Society for the Promotion of Science (JSPS) KAKENHI and Japanese Ministry of Education, Culture, Sports, Science and Technology (MEXT), Japan;
Consejo Nacional de Ciencia (CONACYT) y Tecnolog\'{i}a, through Fondo de Cooperaci\'{o}n Internacional en Ciencia y Tecnolog\'{i}a (FONCICYT) and Direcci\'{o}n General de Asuntos del Personal Academico (DGAPA), Mexico;
Nederlandse Organisatie voor Wetenschappelijk Onderzoek (NWO), Netherlands;
The Research Council of Norway, Norway;
Commission on Science and Technology for Sustainable Development in the South (COMSATS), Pakistan;
Pontificia Universidad Cat\'{o}lica del Per\'{u}, Peru;
Ministry of Science and Higher Education and National Science Centre, Poland;
Korea Institute of Science and Technology Information and National Research Foundation of Korea (NRF), Republic of Korea;
Ministry of Education and Scientific Research, Institute of Atomic Physics and Romanian National Agency for Science, Technology and Innovation, Romania;
Joint Institute for Nuclear Research (JINR), Ministry of Education and Science of the Russian Federation and National Research Centre Kurchatov Institute, Russia;
Ministry of Education, Science, Research and Sport of the Slovak Republic, Slovakia;
National Research Foundation of South Africa, South Africa;
Centro de Aplicaciones Tecnol\'{o}gicas y Desarrollo Nuclear (CEADEN), Cubaenerg\'{\i}a, Cuba, Ministerio de Ciencia e Innovacion and Centro de Investigaciones Energ\'{e}ticas, Medioambientales y Tecnol\'{o}gicas (CIEMAT), Spain;
Swedish Research Council (VR) and Knut \& Alice Wallenberg Foundation (KAW), Sweden;
European Organization for Nuclear Research, Switzerland;
National Science and Technology Development Agency (NSDTA), Suranaree University of Technology (SUT) and Office of the Higher Education Commission under NRU project of Thailand, Thailand;
Turkish Atomic Energy Agency (TAEK), Turkey;
National Academy of  Sciences of Ukraine, Ukraine;
Science and Technology Facilities Council (STFC), United Kingdom;
National Science Foundation of the United States of America (NSF) and United States Department of Energy, Office of Nuclear Physics (DOE NP), United States of America.

%% file: Alice_Authorlist_2017-Apr-05.tex


\begingroup
\small
\begin{flushleft}
S.~Acharya$^\textrm{\scriptsize 139}$,
D.~Adamov\'{a}$^\textrm{\scriptsize 96}$,
J.~Adolfsson$^\textrm{\scriptsize 34}$,
M.M.~Aggarwal$^\textrm{\scriptsize 101}$,
G.~Aglieri Rinella$^\textrm{\scriptsize 35}$,
M.~Agnello$^\textrm{\scriptsize 31}$,
N.~Agrawal$^\textrm{\scriptsize 48}$,
Z.~Ahammed$^\textrm{\scriptsize 139}$,
N.~Ahmad$^\textrm{\scriptsize 17}$,
S.U.~Ahn$^\textrm{\scriptsize 80}$
% ...
\renewcommand\labelenumi{\textsuperscript{\theenumi}~}

\section*{Affiliation notes}
\renewcommand\theenumi{\roman{enumi}}
\begin{Authlist}
\item \Adef{0}Deceased
\item \Adef{idp1835040}{Also at: Dipartimento DET del Politecnico di Torino, Turin, Italy}
\item \Adef{idp1854432}{Also at: Georgia State University, Atlanta, Georgia, United States}
\item \Adef{idp4144160}{Also at: M.V. Lomonosov Moscow State University, D.V. Skobeltsyn Institute of Nuclear, Physics, Moscow, Russia}
\item \Adef{idp4502608}{Also at: Department of Applied Physics, Aligarh Muslim University, Aligarh, India}
\item \Adef{idp5486048}{Also at: Institute of Theoretical Physics, University of Wroclaw, Poland}
\end{Authlist}

\section*{Collaboration Institutes}
\renewcommand\theenumi{\arabic{enumi}~}

$^{1}$A.I. Alikhanyan National Science Laboratory (Yerevan Physics Institute) Foundation, Yerevan, Armenia
\\
$^{2}$Benem\'{e}rita Universidad Aut\'{o}noma de Puebla, Puebla, Mexico
\\
$^{3}$Bogolyubov Institute for Theoretical Physics, Kiev, Ukraine
\\
$^{4}$Bose Institute, Department of Physics 
and Centre for Astroparticle Physics and Space Science (CAPSS), Kolkata, India
\\
$^{5}$Budker Institute for Nuclear Physics, Novosibirsk, Russia
\\
$^{6}$California Polytechnic State University, San Luis Obispo, California, United States
\\
$^{7}$Central China Normal University, Wuhan, China
\\
$^{8}$Centre de Calcul de l'IN2P3, Villeurbanne, Lyon, France
\\
$^{9}$Centro de Aplicaciones Tecnol\'{o}gicas y Desarrollo Nuclear (CEADEN), Havana, Cuba
\\
$^{10}$Centro de Investigaciones Energ\'{e}ticas Medioambientales y Tecnol\'{o}gicas (CIEMAT), Madrid, Spain
\\
$^{11}$Centro de Investigaci\'{o}n y de Estudios Avanzados (CINVESTAV), Mexico City and M\'{e}rida, Mexico
\\
$^{12}$Centro Fermi - Museo Storico della Fisica e Centro Studi e Ricerche ``Enrico Fermi', Rome, Italy
\\
$^{13}$Chicago State University, Chicago, Illinois, United States
\\
$^{14}$China Institute of Atomic Energy, Beijing, China
\\
$^{15}$COMSATS Institute of Information Technology (CIIT), Islamabad, Pakistan
\\
$^{16}$Departamento de F\'{\i}sica de Part\'{\i}culas and IGFAE, Universidad de Santiago de Compostela, Santiago de Compostela, Spain
\\
$^{17}$Department of Physics, Aligarh Muslim University, Aligarh, India
\\
$^{18}$Department of Physics, Ohio State University, Columbus, Ohio, United States
\\
$^{19}$Department of Physics, Pusan National University, Pusan, South Korea
\\
$^{20}$Department of Physics, Sejong University, Seoul, South Korea
\\
$^{21}$Department of Physics, University of Oslo, Oslo, Norway
\\
$^{22}$Department of Physics and Technology, University of Bergen, Bergen, Norway
\\
$^{23}$Dipartimento di Fisica dell'Universit\`{a} 'La Sapienza'
and Sezione INFN, Rome, Italy
\\
$^{24}$Dipartimento di Fisica dell'Universit\`{a}
and Sezione INFN, Cagliari, Italy
\\
$^{25}$Dipartimento di Fisica dell'Universit\`{a}
and Sezione INFN, Trieste, Italy
\\
$^{26}$Dipartimento di Fisica dell'Universit\`{a}
and Sezione INFN, Turin, Italy
\\
$^{27}$Dipartimento di Fisica e Astronomia dell'Universit\`{a}
and Sezione INFN, Bologna, Italy
\\
$^{28}$Dipartimento di Fisica e Astronomia dell'Universit\`{a}
and Sezione INFN, Catania, Italy
\\
$^{29}$Dipartimento di Fisica e Astronomia dell'Universit\`{a}
and Sezione INFN, Padova, Italy
\\
$^{30}$Dipartimento di Fisica `E.R.~Caianiello' dell'Universit\`{a}
and Gruppo Collegato INFN, Salerno, Italy
\\
$^{31}$Dipartimento DISAT del Politecnico and Sezione INFN, Turin, Italy
\\
$^{32}$Dipartimento di Scienze e Innovazione Tecnologica dell'Universit\`{a} del Piemonte Orientale and INFN Sezione di Torino, Alessandria, Italy
\\
$^{33}$Dipartimento Interateneo di Fisica `M.~Merlin'
and Sezione INFN, Bari, Italy
\\
$^{34}$Division of Experimental High Energy Physics, University of Lund, Lund, Sweden
\\
$^{35}$European Organization for Nuclear Research (CERN), Geneva, Switzerland
\\
$^{36}$Excellence Cluster Universe, Technische Universit\"{a}t M\"{u}nchen, Munich, Germany
\\
$^{37}$Faculty of Engineering, Bergen University College, Bergen, Norway
\\
$^{38}$Faculty of Mathematics, Physics and Informatics, Comenius University, Bratislava, Slovakia
\\
$^{39}$Faculty of Nuclear Sciences and Physical Engineering, Czech Technical University in Prague, Prague, Czech Republic
\\
$^{40}$Faculty of Science, P.J.~\v{S}af\'{a}rik University, Ko\v{s}ice, Slovakia
\\
$^{41}$Faculty of Technology, Buskerud and Vestfold University College, Tonsberg, Norway
\\
$^{42}$Frankfurt Institute for Advanced Studies, Johann Wolfgang Goethe-Universit\"{a}t Frankfurt, Frankfurt, Germany
\\
$^{43}$Gangneung-Wonju National University, Gangneung, South Korea
\\
$^{44}$Gauhati University, Department of Physics, Guwahati, India
\\
$^{45}$Helmholtz-Institut f\"{u}r Strahlen- und Kernphysik, Rheinische Friedrich-Wilhelms-Universit\"{a}t Bonn, Bonn, Germany
\\
$^{46}$Helsinki Institute of Physics (HIP), Helsinki, Finland
\\
$^{47}$Hiroshima University, Hiroshima, Japan
\\
$^{48}$Indian Institute of Technology Bombay (IIT), Mumbai, India
\\
$^{49}$Indian Institute of Technology Indore, Indore, India
\\
$^{50}$Indonesian Institute of Sciences, Jakarta, Indonesia
\\
$^{51}$INFN, Laboratori Nazionali di Frascati, Frascati, Italy
\\
$^{52}$INFN, Laboratori Nazionali di Legnaro, Legnaro, Italy
\\
$^{53}$INFN, Sezione di Bari, Bari, Italy
\\
$^{54}$INFN, Sezione di Bologna, Bologna, Italy
\\
$^{55}$INFN, Sezione di Cagliari, Cagliari, Italy
\\
$^{56}$INFN, Sezione di Catania, Catania, Italy
\\
$^{57}$INFN, Sezione di Padova, Padova, Italy
\\
$^{58}$INFN, Sezione di Roma, Rome, Italy
\\
$^{59}$INFN, Sezione di Torino, Turin, Italy
\\
$^{60}$INFN, Sezione di Trieste, Trieste, Italy
\\
$^{61}$Inha University, Incheon, South Korea
\\
$^{62}$Institut de Physique Nucl\'eaire d'Orsay (IPNO), Universit\'e Paris-Sud, CNRS-IN2P3, Orsay, France
\\
$^{63}$Institute for Nuclear Research, Academy of Sciences, Moscow, Russia
\\
$^{64}$Institute for Subatomic Physics of Utrecht University, Utrecht, Netherlands
\\
$^{65}$Institute for Theoretical and Experimental Physics, Moscow, Russia
\\
$^{66}$Institute of Experimental Physics, Slovak Academy of Sciences, Ko\v{s}ice, Slovakia
\\
$^{67}$Institute of Physics, Academy of Sciences of the Czech Republic, Prague, Czech Republic
\\
$^{68}$Institute of Physics, Bhubaneswar, India
\\
$^{69}$Institute of Space Science (ISS), Bucharest, Romania
\\
$^{70}$Institut f\"{u}r Informatik, Johann Wolfgang Goethe-Universit\"{a}t Frankfurt, Frankfurt, Germany
\\
$^{71}$Institut f\"{u}r Kernphysik, Johann Wolfgang Goethe-Universit\"{a}t Frankfurt, Frankfurt, Germany
\\
$^{72}$Institut f\"{u}r Kernphysik, Westf\"{a}lische Wilhelms-Universit\"{a}t M\"{u}nster, M\"{u}nster, Germany
\\
$^{73}$Instituto de Ciencias Nucleares, Universidad Nacional Aut\'{o}noma de M\'{e}xico, Mexico City, Mexico
\\
$^{74}$Instituto de F\'{i}sica, Universidade Federal do Rio Grande do Sul (UFRGS), Porto Alegre, Brazil
\\
$^{75}$Instituto de F\'{\i}sica, Universidad Nacional Aut\'{o}noma de M\'{e}xico, Mexico City, Mexico
\\
$^{76}$IRFU, CEA, Universit\'{e} Paris-Saclay, Saclay, France
\\
$^{77}$iThemba LABS, National Research Foundation, Somerset West, South Africa
\\
$^{78}$Joint Institute for Nuclear Research (JINR), Dubna, Russia
\\
$^{79}$Konkuk University, Seoul, South Korea
\\
$^{80}$Korea Institute of Science and Technology Information, Daejeon, South Korea
\\
$^{81}$KTO Karatay University, Konya, Turkey
\\
$^{82}$Laboratoire de Physique Corpusculaire (LPC), Clermont Universit\'{e}, Universit\'{e} Blaise Pascal, CNRS--IN2P3, Clermont-Ferrand, France
\\
$^{83}$Laboratoire de Physique Subatomique et de Cosmologie, Universit\'{e} Grenoble-Alpes, CNRS-IN2P3, Grenoble, France
\\
$^{84}$Lawrence Berkeley National Laboratory, Berkeley, California, United States
\\
$^{85}$Moscow Engineering Physics Institute, Moscow, Russia
\\
$^{86}$Nagasaki Institute of Applied Science, Nagasaki, Japan
\\
$^{87}$National and Kapodistrian University of Athens, Physics Department, Athens, Greece, Athens, Greece
\\
$^{88}$National Centre for Nuclear Studies, Warsaw, Poland
\\
$^{89}$National Institute for Physics and Nuclear Engineering, Bucharest, Romania
\\
$^{90}$National Institute of Science Education and Research, Bhubaneswar, India
\\
$^{91}$National Nuclear Research Center, Baku, Azerbaijan
\\
$^{92}$National Research Centre Kurchatov Institute, Moscow, Russia
\\
$^{93}$Niels Bohr Institute, University of Copenhagen, Copenhagen, Denmark
\\
$^{94}$Nikhef, Nationaal instituut voor subatomaire fysica, Amsterdam, Netherlands
\\
$^{95}$Nuclear Physics Group, STFC Daresbury Laboratory, Daresbury, United Kingdom
\\
$^{96}$Nuclear Physics Institute, Academy of Sciences of the Czech Republic, \v{R}e\v{z} u Prahy, Czech Republic
\\
$^{97}$Oak Ridge National Laboratory, Oak Ridge, Tennessee, United States
\\
$^{98}$Petersburg Nuclear Physics Institute, Gatchina, Russia
\\
$^{99}$Physics Department, Creighton University, Omaha, Nebraska, United States
\\
$^{100}$Physics department, Faculty of science, University of Zagreb, Zagreb, Croatia
\\
$^{101}$Physics Department, Panjab University, Chandigarh, India
\\
$^{102}$Physics Department, University of Cape Town, Cape Town, South Africa
\\
$^{103}$Physics Department, University of Jammu, Jammu, India
\\
$^{104}$Physics Department, University of Rajasthan, Jaipur, India
\\
$^{105}$Physikalisches Institut, Eberhard Karls Universit\"{a}t T\"{u}bingen, T\"{u}bingen, Germany
\\
$^{106}$Physikalisches Institut, Ruprecht-Karls-Universit\"{a}t Heidelberg, Heidelberg, Germany
\\
$^{107}$Physik Department, Technische Universit\"{a}t M\"{u}nchen, Munich, Germany
\\
$^{108}$Purdue University, West Lafayette, Indiana, United States
\\
$^{109}$Research Division and ExtreMe Matter Institute EMMI, GSI Helmholtzzentrum f\"ur Schwerionenforschung GmbH, Darmstadt, Germany
\\
$^{110}$Rudjer Bo\v{s}kovi\'{c} Institute, Zagreb, Croatia
\\
$^{111}$Russian Federal Nuclear Center (VNIIEF), Sarov, Russia
\\
$^{112}$Saha Institute of Nuclear Physics, Kolkata, India
\\
$^{113}$School of Physics and Astronomy, University of Birmingham, Birmingham, United Kingdom
\\
$^{114}$Secci\'{o}n F\'{\i}sica, Departamento de Ciencias, Pontificia Universidad Cat\'{o}lica del Per\'{u}, Lima, Peru
\\
$^{115}$SSC IHEP of NRC Kurchatov institute, Protvino, Russia
\\
$^{116}$Stefan Meyer Institut f\"{u}r Subatomare Physik (SMI), Vienna, Austria
\\
$^{117}$SUBATECH, IMT Atlantique, Universit\'{e} de Nantes, CNRS-IN2P3, Nantes, France
\\
$^{118}$Suranaree University of Technology, Nakhon Ratchasima, Thailand
\\
$^{119}$Technical University of Ko\v{s}ice, Ko\v{s}ice, Slovakia
\\
$^{120}$Technical University of Split FESB, Split, Croatia
\\
$^{121}$The Henryk Niewodniczanski Institute of Nuclear Physics, Polish Academy of Sciences, Cracow, Poland
\\
$^{122}$The University of Texas at Austin, Physics Department, Austin, Texas, United States
\\
$^{123}$Universidad Aut\'{o}noma de Sinaloa, Culiac\'{a}n, Mexico
\\
$^{124}$Universidade de S\~{a}o Paulo (USP), S\~{a}o Paulo, Brazil
\\
$^{125}$Universidade Estadual de Campinas (UNICAMP), Campinas, Brazil
\\
$^{126}$Universidade Federal do ABC, Santo Andre, Brazil
\\
$^{127}$University of Houston, Houston, Texas, United States
\\
$^{128}$University of Jyv\"{a}skyl\"{a}, Jyv\"{a}skyl\"{a}, Finland
\\
$^{129}$University of Liverpool, Liverpool, United Kingdom
\\
$^{130}$University of Tennessee, Knoxville, Tennessee, United States
\\
$^{131}$University of the Witwatersrand, Johannesburg, South Africa
\\
$^{132}$University of Tokyo, Tokyo, Japan
\\
$^{133}$University of Tsukuba, Tsukuba, Japan
\\
$^{134}$Universit\'{e} de Lyon, Universit\'{e} Lyon 1, CNRS/IN2P3, IPN-Lyon, Villeurbanne, Lyon, France
\\
$^{135}$Universit\'{e} de Strasbourg, CNRS, IPHC UMR 7178, F-67000 Strasbourg, France, Strasbourg, France
\\
$^{136}$Universit\`{a} degli Studi di Pavia, Pavia, Italy
\\
$^{137}$Universit\`{a} di Brescia, Brescia, Italy
\\
$^{138}$V.~Fock Institute for Physics, St. Petersburg State University, St. Petersburg, Russia
\\
$^{139}$Variable Energy Cyclotron Centre, Kolkata, India
\\
$^{140}$Warsaw University of Technology, Warsaw, Poland
\\
$^{141}$Wayne State University, Detroit, Michigan, United States
\\
$^{142}$Wigner Research Centre for Physics, Hungarian Academy of Sciences, Budapest, Hungary
\\
$^{143}$Yale University, New Haven, Connecticut, United States
\\
$^{144}$Yonsei University, Seoul, South Korea
\\
$^{145}$Zentrum f\"{u}r Technologietransfer und Telekommunikation (ZTT), Fachhochschule Worms, Worms, Germany
\endgroup